\documentclass[12pt,showpacs,preprint,nofootinbib,showkeys]{revtex4}

\usepackage[sumlimits,intlimits,namelimits]{amsmath}
\usepackage{bbm}
\usepackage{epsfig}

\renewcommand{\vec}[1]{\boldsymbol{#1}}

\newcommand{\im}{\text{i}}
\newcommand{\ud}{\text{d}}
\newcommand{\h}{\frac{1}{2}}
\newcommand{\lk}{\left(}
\newcommand{\rk}{\right)}
\newcommand{\bR}{\mathbbm{R}}

\newcommand{\bZ}{\mathbbm{Z}}

\begin{document}

\setlength{\parindent}{0cm}

\vspace{2cm}

\title{Energy Density of Vortices in the Schr\"odinger Picture}

\hfill{FSU TPI 02/03}

\author{J.D. L\"ange\footnotemark[1], M. Engelhardt\footnotemark[2]
and H. Reinhardt\footnotemark[2]}

\vspace{1cm}

\affiliation{
\footnotemark[1] Theoretisch-Physikalisches Institut,
Friedrich-Schiller-Universit\"at Jena, Fr\"obelstieg 1, 07743 Jena,
Germany \footnotetext[1]{\tt J.D.Laenge@tpi.uni-jena.de} \\
\footnotemark[2] Institut f\"ur Theoretische Physik,
Eberhard-Karls-Universit\"at T\"ubingen, Auf der Morgenstelle 14, 72076
T\"ubingen, Germany \footnotetext[2]{\tt
engelm@tphys.physik.uni-tuebingen.de,
hugo.reinhardt@uni-tuebingen.de} }

\vspace{1cm}

\begin{abstract}
\noindent
The one-loop
energy density of an infinitely thin static magnetic vortex in $SU (2)$
Yang-Mills theory is evaluated using the Schr\"odinger picture. Both the
gluonic fluctuations as well as the quarks in the vortex background are
included. The
energy density of the magnetic vortex is discussed as
a function of the magnetic flux. The center vortices correspond to local minima
in the effective potential. These minima are degenerated with the perturbative
vacuum if the fermions are ignored. Inclusion of fermions lifts this
degeneracy, raising the vortex energy above the energy of the perturbative
vacuum.
\end{abstract}

\pacs{02.30.Sa, 02.30.Tb, 03.70.+k, 11.15.-q}

\keywords{center vortices, self-adjoint extensions, one-loop energy}

\maketitle

\section{Introduction}

Recently, there has been a revival of the vortex picture of confinement
\cite{R1}. Lattice calculations performed in the so-called maximal center
gauge with subsequent center projection \cite{R2} (which provides a simple
method for identifying the center vortices in Yang-Mills configurations) have
given substantial support to the idea that center vortices are the relevant
infrared degrees of freedom responsible for confinement. The deconfinement
phase transition in the vortex picture takes the guise of a percolation
transition \cite{R5,rand}. Furthermore, the topological charge of the vortices
can be understood in terms of their self-intersection number
\cite{R6,preptop,moretop, RX1, RX2} and the spontaneous breaking of chiral symmetry
can be generated through vortex effects \cite{forc, csb}.

On the other hand, while a successful phenomenological model of vortex
dynamics has been developed \cite{rand,preptop,csb}, 
investigations of the dynamical properties of Yang-Mills vortices have
only recently been launched, cf. \cite{maul,bordag} for thick center
vortices and \cite{R3} for thin vortices of arbitrary flux. In the present
paper, we expand on these investigations, which only considered gluonic
fluctuations to one-loop order, by evaluating the one-loop energy density
of an infinitely thin static magnetic vortex including quark fluctuations.
For this purpose we use the Schr\"odinger picture.  We
solve the Schr\"odinger equation for gluonic fluctuations and fermions
moving in the vortex background field.
{From} the obtained eigenenergies we calculate the
static energy density of the vortex as a function of its magnetic flux.
We will show that the
inclusion of fermions substantially changes the energy density of the vortex.
In particular we will find that the fermions raise the
center vortex effective energy compared with the perturbative vacuum.

The organization of the paper is as follows: In the next section we study the
Schr\"odinger equation for the gluonic fluctuations around the vortex
background field.
In subsection \ref{subsec21} we study the Hamiltonian of the gluonic
fluctuations in the vortex background and discuss its self-adjoint extensions
in subsection \ref{subsec22}. Subsection \ref{subsec23} is devoted to the
incorporation of Gauss' law and to the selection of the appropriate
self-adjoint extension of the Hamiltonian of the gluonic fluctuations.
In subsection \ref{subsec24} the
gluonic contribution to the energy density of the vortex is presented.

In section \ref{sec3} we study fermions in the vortex background field and
calculate their contribution to the energy density. The physical
interpretation of our results and some concluding remarks are given in
section \ref{sec4}.

\section{Gluonic fluctuations around the vortex background field}

\subsection{The Hamiltonian of the gluonic fluctuations}
\label{subsec21}

Our aim is to evaluate the one-loop corrections to the energy density of a
static SU(2) vortex background field. We consider an infinitely thin
vortex of magnetic flux $\Phi$ on the 3-axis described by a gauge potential

\begin{eqnarray}
\label{G1}\vec{\bar{A}} (\rho,\varphi )=
\frac{\sigma^3}{2} \frac{\Phi}{g \rho}
\hat{\vec{e}}_{\varphi}\;,
\end{eqnarray}
where the polar coordinates $(\rho,\varphi )$ parameterize the 
$(1,2)$-plane in space. The normalization has been chosen such that odd 
integer values of $\Phi $ describe a center vortex, i.e. any Wilson
loop $W$ in the $(1,2)$-plane encircling the origin takes the value
$W=-1$. The magnetic field associated with (\ref{G1}) is

\begin{equation}
\vec{\bar{B}}^a (\vec{x})=\delta^{a3} 
\frac{2\pi\Phi }{g} \delta(x_1)\delta(x_2)\hat{\vec{e}}_z\;.
\label{G2}
\end{equation}
Our starting point for the evaluation of the one-loop energy density is the
Yang-Mills Hamiltonian in Weyl gauge $(A_0 = 0)$ 
in the presence of an external
current
\begin{equation}
H=\int \ud^3 x \, \left[ \frac{1}{2} E^a_i E^a_i
+ \frac{1}{2} B^a_i B^a_i - j^a_i A^a_i \right] \;.
\end{equation}
By decomposing the gauge field $A^a_i$ 
into the vortex background field $\bar{A}^a_i$ and fluctuations $\eta^a_i$,
\begin{equation}
A^a_i = \bar{A}^{a}_{i} +\eta^{a}_{i} \;,
\end{equation}
dropping terms of higher order than quadratic in the fluctuations $\eta $
(one-loop approximation), adjusting the external current such as to induce
the background field $\vec{\bar{A}}^a $ at the classical level, i.e.
\begin{equation}
\bar{D}^{ab}_{i} \bar{F}^{b}_{ij} = -j_j^a \;, \ \
\bar{D}^{ab}_{i} = \delta^{ab} \partial_{i} + gf^{abc} \bar{A}^{c}_{i}
\end{equation}
and finally choosing the background gauge
\begin{equation}
\bar{D}^{ab}_{i} \eta_{i}^{b} = 0
\end{equation} 
for the fluctuations, one obtains the Hamiltonian (in the ``coordinate''
representation)

\begin{equation}
\begin{array}{lcl}
H & = & H^{\text{cl}} + H^{(1)} \qquad \text{with} \\
H^{\text{cl}} & = & \h \int \ud^3 x \, \left[
\bar{B}_{i}^{a} \bar{B}_{i}^{a} - 2 j_i^a \bar{A}_{i}^{a} \right] \qquad
\text{and} \\
H^{(1)} & = & \h \int \ud^3 x \, \left[
-\frac{\delta^{2} }{(\delta \eta_{i}^{a} )^2 }
+\eta_{i}^{a} \hat{O}_{ij}^{ab} \eta_{j}^{b} \right].
\end{array}
\label{G3}
\end{equation}
Here the operator $\hat{O} $ reads
\begin{equation}
\hat{O}_{ij}^{ab} = -\delta_{ij} (\bar{D}^{2} )^{ab} 
+ 2gf^{abc} \epsilon_{ijk} \bar{B}_{k}^{c}\;,
\label{opero}
\end{equation}
where $(\bar{D}^2)^{ab} = \bar{D}_i^{ac} \bar{D}_i^{cb}$.
In this approximation, the ground state wave functional takes the
Gaussian form
\begin{equation}
\Psi [\eta ] = \exp \left( -\frac{1}{2} \int \eta K \eta \right)
\end{equation}
with the spectral representation for the covariance $K$
\begin{equation}
K_{ij}^{ab} (\vec{x},\vec{y}) = \sum_{N} (\phi^{\dagger } )^{a}_{i}
(N,\vec{x}) \sqrt{O(N)} \phi^{b}_{j} (N,\vec{y}) .
\end{equation}
Here $\phi (N)$ denotes the eigenfunctions of $\hat{O}$, cf.~(\ref{opero}),
\begin{equation}
\label{13}
\hat{O} \phi (N) = O (N) \phi (N) ,
\end{equation}
where $N$ denotes a (not necessarily enumerable) complete set of
quantum numbers. Furthermore, the eigenfunctions $\phi (N)$ have to fulfill the
``background gauge'' condition
\begin{eqnarray}
\bar{D} \phi (N) &=& 0 \label{G3a} \;,
\end{eqnarray}
such that the wave functional $\Psi $ fulfills Gauss' law,

\begin{equation}
\left(\bar{D}_{i}^{ab} + gf^{abc} \eta_{i}^{c} \right)
\frac{\delta }{\delta \eta_{i}^{b} } \Psi [\eta ] = 0
\label{G4a}
\end{equation} 
to the desired order, namely ${\cal O} (g^0 )$, and is an
eigenfunctional of the Hamiltonian (\ref{G3}) with eigenvalue
\begin{equation}
E = E^{\text{cl}} + \frac{1}{2} \int \ud^3 x \, \underbrace{\left[
\sum_{N} (\phi^{\dagger } )_{i}^{a} (N,x) 
\sqrt{O(N)} \phi_{i}^{a} (N,x) \right]}_{K^{a a}_{i i} (x, x)}
\equiv \int \ud^3 x \, e(x) \;,
\end{equation}
where $e(x)$ denotes the energy density.
To calculate $e(x)$, the set of eigenfunctions $\phi $
must be specified. As a first step, it is useful to reduce the problem by
decomposing the functions $\phi $ (and the quantum numbers $N$) as

\begin{equation}
\phi_{j}^{b} (N,x) = u_j (i) e^b (a) f_M ( x_1 , x_2 ) \exp (\im kx_3)
\exp(\im m \varphi)
\;, \qquad k\in \mathbbm{R},\; m \in \mathbbm{Z}
\label{G5}
\end{equation}
with polar color and Lorentz unit vectors

\begin{equation}
\begin{array}{lll}
e(1) = (1/\sqrt{2} ,\im/\sqrt{2} ,0)\;, &
e(2) = (1/\sqrt{2} ,-\im/\sqrt{2} ,0)\;, & e(3) = (0,0,1)\;, \\
u(1) = (1/\sqrt{2} ,\im/\sqrt{2} ,0)\;, &
u(2) = (1/\sqrt{2} ,-\im/\sqrt{2} ,0)\;, & u(3) = (0,0,1)\;,
\end{array}
\label{G6}
\end{equation}
in terms of which the eigenvalue equation (\ref{13}) reduces to

\begin{equation}
\hat{B} (a,i,m) f_M = B(M,a,i,m) f_M
\label{G6a}
\end{equation}
where the operator $\hat{B}$ is defined by 

\begin{equation}
\hat{B} (a, i, m) = -\frac{1}{\rho} \frac{\ud}{\ud\rho} \rho
\frac{\ud}{\ud\rho}
+ \frac{(m-\lambda(a) \Phi)^2}{\rho^2} -2 \lambda(i) \lambda(a)
2 \pi \Phi \delta(x_1) \delta(x_2)
\label{G7}
\end{equation}
and its eigenvalues $B (M, a, i, m)$ are related to the ones of
$\hat{O}$, cf.~(\ref{opero}), by
\begin{equation}
O (N) = B (M, a, i, m) + k^2 \;.
\end{equation}
Note that the color quantum numbers $a$ and the Lorentz quantum numbers
$i$ enter via the eigenvalues $\lambda $ corresponding to the
vectors (\ref{G6}), namely
\begin{equation}
\lambda (1) = 1 \qquad \lambda (2) = -1 \qquad \lambda (3) = 0 \;.
\end{equation}
The additional constraint on the $f_M$ imposed by 
Gauss' law, eq. (\ref{G4a}), will be dealt with later.
As the operator $\hat{B}$, cf.~(\ref{G7}), is not well defined on the
$z$-axis, we restrict space to $\mathbbm{R}^3\backslash \{z-\text{axis}\}$
and take the delta functions into account by adopting
appropriate boundary conditions.

At this stage, it is useful to introduce a scale $\bar{M} $ which has the
dimension of a mass. This allows us to make 
eigenvalue equations dimensionless and, when
we determine boundary conditions, we make an expansion of functions
for distances to the $z$-axis small compared to the scale
$1/\bar{M} $. With this scale, we define a new variable
$x:=\rho \bar{M} $ and obtain a dimensionless operator
\begin{eqnarray}
\hat{C} & = & \frac{1}{\bar{M}^{2} } \hat{B} = - \frac{1}{x}
\frac{\ud}{\ud x} x \frac{\ud}{\ud x} +
\frac{(m-\lambda(a)\Phi)^2}{x^2}\;. \label{chat}
\end{eqnarray}
For a complete definition of the set of functions $f_M$, a set of boundary
conditions in $L^2(\mathbbm{R}_{>0}, x \ud x)$ must be
specified for $x \rightarrow 0$, such that $\hat{C}$ is
self-adjoint.
In order to systematically construct all self-adjoint
extensions, we initially restrict the domain $D(\hat{C})$ to $D(\hat{C}) =
C^\infty_c(\mathbbm{R}_{>0})$ 
on which $\hat{C} $ is a well defined symmetric operator, i.e.
\begin{equation}
\langle f | \hat{C} g \rangle = \langle \hat{C} f | g \rangle \qquad 
\mbox{for all} \ f,g \in D(\hat{C} ) \ .
\end{equation}
Self-adjointness, by contrast, means more; namely, that $\hat{C} $ is
equal to its adjoint $\hat{C}^{\dagger } $, defined as follows. Let
$D(\hat{C}^{\dagger } )$ be the set of functions $g$ for which $h \in
L^2(\mathbbm{R}_{>0}, x \ud x)$ exists such that

\begin{equation}
\langle g | \hat{C} f \rangle = \langle h | f \rangle \qquad
\mbox{for all} \ f \in D(\hat{C} )\;.
\label{G8}
\end{equation}
Then, for each $g \in D(\hat{C}^{\dagger } )$, one defines 
$\hat{C}^{\dagger } g = h$. As $D(\hat{C}) =
C^\infty_c(\mathbbm{R}_{>0})$, one sees that in the sense of weak
derivatives $-\frac{1}{x} \frac{\ud}{\ud x} x \frac{\ud}{\ud x} g$ is
locally in $L^2(\mathbbm{R}_{>0},x \ud x)$ and therefore
also locally in $L^1(\mathbbm{R}_{>0},x \ud x)$. From this one can conclude
that $(xg')'$ is locally in $L^2(\mathbbm{R}_{>0} , \ud x)$ with the usual
measure and, with Sobolev's
lemma \cite{Reed}, that $g'$ is absolute
continuous. This allows an integration by parts in (\ref{G8}) which
gives

\begin{eqnarray}
\label{G9a}
\hat{C}^\dagger & = & -\frac{1}{x} \frac{\ud}{\ud x} x \frac{\ud}{\ud x} +
\frac{\delta^2}{x^2} \\
D(\hat{C}^\dagger) & = & \{ f \in L^2(\mathbbm{R}_{>0} , x \ud x) | f'
\text{ absolute continuous, }\hat{C}^\dagger f \in
L^2(\mathbbm{R}_{>0},x \ud x) \}\;,
\label{G9b}
\end{eqnarray}
where the derivatives are now ordinary derivatives and we introduced
$\delta = | m - \lambda(a) \Phi|$.
Note that $D(\hat{C} ) \subset D(\hat{C}^{\dagger } )$ as it should be
for the symmetric operator $\hat{C}$. In the following, we construct all
self-adjoint extensions of $\hat{C}$. In the cases where we obtain more
than one self-adjoint extension, which corresponds to different possible
boundary conditions on the $z$-axis, the choice of a definite self-adjoint
extension is a question of the physical situation one wishes to describe;
mathematically, all choices are equally valid.

\subsection{Self-adjoint extensions}
\label{subsec22}
To determine
self-adjoint extensions of $\hat{C} $, it is necessary to consider the
deficiency spaces $\mathcal{H}_{\pm } $ of $\hat{C} $, i.e. the spaces of
zero modes of
the operators $\hat{C}^{\dagger} \mp \im $, with their respective dimensions
(deficiency indices) $n_{\pm } $ \cite{Reed}. The operator $\hat{C}$ is
called essentially self-adjoint if $n_{+} = n_{-} =0$, thus admitting a
unique self-adjoint extension given by its closure.
With $f \in D(\hat{C}^{\dagger} )$, cf.~(\ref{G9b}), and
\begin{eqnarray}
\hat{C}^\dagger f & = & \pm \im f
\end{eqnarray}
one can conclude that
$f \in C^\infty_c(\mathbbm{R}_{>0})$. Two linearly independent solutions
of $\hat{C^\dagger} f=\im f$ are
$I_\delta\left(\frac{1}{\sqrt{2}}(1-i)x \right)$ and
$K_\delta\left(\frac{1}{\sqrt{2}}(1-i)x \right)$, where $K_\delta$ and
$I_\delta$ denote  the modified
Bessel functions. With the condition $f \in D(\hat{C}^{\dagger} )$ only the
solution 
\begin{equation}
f_+(x) = K_\delta \left(\frac{1}{\sqrt{2}}(1-i)x \right)
\end{equation}
 for
$0 \leq \delta < 1$ remains. Therefore
\begin{equation}
\begin{array}{lcll}
\mathcal{H}_+ & = & \{ cf_+ | c \in \mathbbm{C} \} & \text{for } 0
\leq \delta < 1 \;, \\
\mathcal{H}_+ & = & \emptyset & \text{for } \delta \geq 1\;.
\end{array}
\end{equation}
Correspondingly, we obtain
\begin{equation}
\begin{array}{lcll}
\mathcal{H}_- & = & \left\{cf_-|c\in\mathbbm{C}\right\} & \text{for }
0 \leq \delta <1 \,,\\
\mathcal{H}_- & = & \emptyset & \text{for } \delta \geq 1\;
\end{array}
\end{equation}
with 
\begin{equation}
f_- ( x ) = K_\delta \left (\frac{1}{\sqrt{2}} (1+ \im) x
\right). 
\end{equation}
As already discussed above, this means that there
exists a unique self-adjoint extension for $\delta \geq 1$. 

For $0 \leq
\delta <1$ the deficiency indices are equal: $n_+ = n_- = 1$, such that  there
is a one parameter family
of self-adjoint extensions
$\hat{C}_\alpha$ ($\alpha \in (-\pi , \pi]$) which are in one-to-one
correspondence to isometries $U_\alpha: f_+ \mapsto \exp(\im \alpha)
f_-$ from $\mathcal{H}_+$ to $\mathcal{H}_-$ in the following
way:
\begin{eqnarray*}
D(\hat{C}_\alpha) & = & \{ f + c(f_+ + U_\alpha f_+) | f \in D(\bar{\hat{C}}),\; c
\in \mathbbm{C} \}\\
\hat{C}_\alpha (f + c(f_+ + U_\alpha f_+)) & = & \hat{C} f + c ( \im f_+ - \im
U_\alpha f_+)\;.
\end{eqnarray*}
In Appendix \ref{append1}, it is shown that the functions $f \in
D(\overline{\hat{C}})$ in the closure
of $\hat{C}$ are absolute continuous close to the origin and fulfill
$\lim_{x\rightarrow 0}
x^{-\delta} f(x) = 0 $. As the function $f_+$ is singular at the origin,
the boundary conditions at $x=0$ is determined by the behavior of this
function. In the following, we determine these boundary conditions for the
cases  $0 < \delta <1$ and $\delta=0$ separately.
\begin{itemize}
\item $0 < \delta = | m - \lambda(a)| < 1$\\
Expanding $f \in D(\hat{C}_\alpha)$ for small $x$, one obtains

\begin{eqnarray}
f(x) & = & A\lk x^\delta + \beta(\alpha) x^{-\delta} \rk +
\text{o}(x^{\delta})\;, \text{ as } x \rightarrow 0
\label{G9c}
\end{eqnarray}
for a constant $A \in \mathbbm{C}$ and with the one-to-one map
\begin{eqnarray*}
\beta(\alpha) & := & \frac{2^{2\delta}\Gamma(\delta)}{\Gamma(-\delta)}
\frac{\cos(\frac{\alpha}{2}-\frac{\pi}{4}\delta)}{\cos(\frac{\alpha}{2}
+\frac{\pi}{4}\delta)}\;.
\end{eqnarray*}
Since $\hat{C} J_\delta(\kappa x )=\kappa^2 J_\delta(\kappa x )$ and
$\hat{C} N_\delta(\kappa x )=\kappa^2 N_\delta(\kappa x)$,
one can make the following ansatz for the scattering states,
\[AJ_\delta(\kappa x )+BN_\delta(\kappa x )\]
and again expand for small $x$. Comparing this with (\ref{G9c})
yields the condition
\[\frac{A}{B}=\frac{\cos(\frac{\alpha}{2}
+\frac{\pi}{4}\delta)}{\kappa^{2\delta}
\cos(\frac{\alpha}{2}-\frac{\pi}{4}\delta)\sin(\delta\pi)}
-\cot(\delta\pi)\,.\]
As bound state, only the solution $K_\delta$ is admissible, since
$\hat{C} K_\delta(\kappa x )=-\kappa^2 K_\delta(\kappa x )$, and
$K_\delta$ is square-integrable. If one expands $K_\delta$ for small
$x $ and again compares with (\ref{G9c}), one notices that
$K_\delta$ with
\[\kappa=\left(\frac{\cos(\frac{\alpha}{2}
+\frac{\pi}{4}\delta)}{\cos(\frac{\alpha}{2}
-\frac{\pi}{4}\delta)}\right)^{\frac{1}{2\delta}}\]
is a possible eigenfunction, if
\[\frac{\cos(\frac{\alpha}{2}
+\frac{\pi}{4}\delta)}{\cos(\frac{\alpha}{2}-\frac{\pi}{4}\delta)}>0\]
is satisfied.
\item $\delta = | m - \lambda(a)| = 0$\\
Expanding $f \in D(\hat{C}_{\alpha } )$ for small $x$, one obtains

\begin{equation}
\label{G10}
f(x) = A \lk \ln \lk \frac{x}{2} \rk + \beta(\alpha)\rk
+ \text{o}(x^0)\;, \text{ as } x \rightarrow 0
\end{equation}
for a constant $A\in\mathbbm{C}$ and with the one-to-one map
\[\beta(\alpha):=\gamma-\frac{\pi}{4}\tan\lk\frac{\alpha}{2}\rk \;,\]
where $\gamma$ is Euler's constant.
For the scattering states, one makes, as above, the ansatz
\[A J_0(\kappa x)+B N_0(\kappa x) \]
which, after expanding for small $x$ and comparing with 
(\ref{G10}), yields the condition
\[\frac{A}{B}=
-\frac{2}{\pi}\ln(\kappa)-\frac{1}{2}\tan\lk\frac{\alpha}{2}\rk\,.\]
The possible bound state is
\[K_0(\kappa x)\,,\]
provided that
\begin{equation}
\kappa=\exp\lk-\frac{\pi}{4}\tan\lk\frac{\alpha}{2}\rk\rk
\label{selbstfortenergie}
\end{equation}
is satisfied; correspondingly, one has the eigenvalue
\[-\kappa^2=-\exp\lk-\frac{\pi}{2}\tan\lk\frac{\alpha}{2}\rk\rk\,.\]
For $\alpha=\pi$ there is no bound state and the solutions are regular
at the origin ($B=0$).
\end{itemize}

\subsection{Gauss' law and choice of self-adjoint extension}
\label{subsec23}

In the previous section, for the cases 
$0\leq\delta \equiv |m-\lambda(a)\Phi|<1$ a whole family of self-adjoint
extensions of the operator $\hat{C} $ was obtained, labeled by one
continuous parameter $\alpha $; each value corresponds to a different
choice of boundary conditions reflecting different
physical situations. In this section, a definite choice for the
parameter $\alpha $ will be motivated.\\
The operator $\hat{B} $, cf.~(\ref{G7}), can be interpreted physically
as the Hamiltonian of a particle with effective angular momentum (grand spin)
$\delta = m-\lambda(a)\Phi$ in a potential 
$-2\lambda(i)\lambda(a)2\pi\Phi\delta(x_1)\delta(x_2)$. Therefore,
a bound state can only arise if the potential term is negative,
i.e. $\lambda(i)\lambda(a)>0$ for positive flux $\Phi > 0$. 
This is the case for
$\lambda(i) = \lambda(a) =1$ and for $\lambda(i) = \lambda(a) =-1$.
It will now be demonstrated that Gauss' law forbids bound states even in
these two cases.

In order to fulfill Gauss' law, it is necessary to restrict the
set of functions $\phi (N)$ in each subspace of definite eigenvalue
$O(N)$ to those linear combinations which satisfy eq.~(\ref{G3a}).
Since the corresponding operator $\bar{D}_i^{ab} $ does not mix
functions $\phi (N)$ with a different color quantum number $a$, such
linear combinations can be constructed for each $a$ separately.
The same is true for the momentum in 3-direction $k$ and the angular
momentum quantum number $m$. Concentrating on the bound state
solutions in the case $\lambda (a)=\lambda (i)=1$, one has, using
the decomposition (\ref{G5}),
\begin{equation}
\bar{D}_{j}^{ab} \phi_{j}^{b} (N) \overset{\text{!}}{=} 0 \Rightarrow 
(\rho \partial_{\rho } -m+\Phi ) f_M =0
\end{equation}
which for the bound state solutions 
$f_M (\rho ) = K_{|m-\Phi |} (\kappa \rho )$ would require
\begin{eqnarray*}
(|m-\Phi|-m+\Phi)K_{|m-\Phi|}(\kappa \rho )
-\kappa \rho K_{|m-\Phi|+1}(\kappa \rho ) & = & 0\,.
\end{eqnarray*}
For this to be true, both coefficients must vanish; in particular,
$\kappa=0$, which contradicts the condition $\kappa>0$. Therefore,
there can be no bound state for $\lambda (a)=\lambda (i)=1$.
The same is true in the case $\lambda (a)=\lambda (i)=-1$.
For $\Phi\in\mathbbm{Z}$, this singles out a definite choice of the
self-adjoint
extension, namely $\alpha =\pi $, since this is the only choice which
yields no bound state. In this case, $B=0$, i.e. the functions are
regular at the origin. It should be noted that for non-integer
flux $\Phi $, in general there are several possibilities with no bound
states. In the following, however, regularity
at the origin ($B=0$) will be demanded even for non-integer flux.

{From} the physical point of view, this outcome is natural. Gauss' law
enforces gauge invariance; gauge invariance in particular implies that
infinitely thin integer fluxes, which are pure gauges
in the theory without quark degrees of freedom, should be unobservable.
Therefore, the gluonic excitation spectrum in the presence of such
a background field should be equivalent to the free case. However,
it should be emphasized that other interpretations, corresponding to
other choices of self-adjoint extension, are possible and should not
a priori be dismissed as invalid; e.g. the thin vortex fluxes
treated here may be used as idealizations of physical, thick fluxes which
are not pure gauges. Then, a different choice of self-adjoint extension
may be useful to mimic the effects of vortex thickness.

\subsection{Energy density of a vortex to order $g^0$}
\label{subsec24}
For the particular choice of self-adjoint extension adopted above,
in the sectors with $\lambda(a)=\pm 1$ the energy density associated
with a vortex differs from  the one in the vacuum by
\begin{eqnarray}
\label{33}
e_{\pm}=2\frac{\bar{M}^{2} }{2\pi}\int\limits^\infty_0\kappa\, \ud\kappa\,
\frac{1}{2\pi}\int\limits^\infty_{-\infty} \ud k
\sum\limits^\infty_{m=-\infty}\sqrt{(\kappa \bar{M} )^{2}+k^2} \left[
\left(J_{|m\mp\Phi|}(\kappa x )\right)^2-\left(J_{|m|}(\kappa
x )\right)^2\right]\,,\label{vortexenergie1}
\end{eqnarray}
(for $x>0$) where the prefactor $2$ comes from the two physical polarizations
allowed by the Gauss'law constraint. In the sector with $\lambda(a)=0$,
the energy density (compared with the one of the vacuum) obviously vanishes.

Let us first consider an integer-valued flux
 $\Phi \in \mathbbm{N}_0,\,\lambda(a)\pm 1$.
For $\Phi\in\mathbbm{N}_0$, one has
\begin{equation}
\sum\limits_{m=-\infty}^\infty(J_{|m|}(\kappa x))^2=1 \hspace{0.5cm} \mbox{and}
\sum\limits_{m=-\infty}^\infty(J_{|m\mp\Phi|}(\kappa x))^2=1 \;.
\end{equation}
Thus, the energy density (\ref{33}) vanishes for all $x>0$ and therefore
\begin{equation}
e(\rho) = 0 \quad \text{for} \quad \rho > 0 \;.
\end{equation} 

The invisibility of thin
integer vortex fluxes $\Phi $, which represent pure gauge backgrounds,
thus becomes manifest.

Let us now consider fluxes  $\Phi\in (0,1)$.
Consider first the energy density (\ref{33}) in the sector given by 
$\lambda(a)=1$: Rewriting the square root in eq. (\ref{33}) by a 
proper-time integral
\begin{equation}
\sqrt{a} = - \lim\limits_{s_{\text{min}} \rightarrow 0}
\int_{s_{\text{min}} }^\infty \ud s\, (\pi s)^{-1/2}
\frac{\ud}{\ud s} \text{e}^{-sa} \hspace{0.5cm},
\end{equation}
where $s_{\text{min} }$ is an (ultraviolet)
proper-time cut-off we find
\begin{eqnarray*}
e_+ &=&
-2 \bar{M}^{2} \int\limits^\infty_{s_{min}}\ud s\,(\pi s)^{-1/2}
\frac{\ud}{\ud s}\frac{1}{2\pi}\int\limits^\infty_0\kappa\, \ud\kappa\,
\frac{1}{2\pi}\int\limits^\infty_{-\infty} \ud k
\sum\limits^\infty_{m=-\infty}\text{e}^{-s((\bar{M} \kappa)^{2} +k^2)}\\
&&\times\left[
\left(J_{|m-\Phi|}(\kappa x )\right)^2
-\left(J_{|m|}(\kappa x )\right)^2\right]
\end{eqnarray*}
The integrals over $k$ and $\kappa$ can be performed by using the relation
\begin{eqnarray}
\int_0^\infty \kappa \; \ud \kappa \; \text{e}^{-s(\bar{M}\kappa)^2}
J_\nu(\kappa x)^2 & = & \frac{1}{2s\bar{M}^2}
\text{e}^{\frac{x^2}{2s\bar{M}^2}}
I_\nu\left(\frac{x^2}{2s\bar{M}^2}\right)\;, \; \nu = |m-\Phi| \text{ or
} \nu = |m|\;,
\end{eqnarray}
cf.~eq.~6.633.2 in \cite{Grad}, yielding
\begin{equation}
e_+
=-2 \bar{M}^{2} \int\limits^\infty_{s_{min}}\ud s\,
(\pi s)^{-1/2}\frac{\ud}{\ud s} \frac{1}{2\pi}\sqrt{\frac{\pi}{s}}
\sum\limits^\infty_{m=-\infty}\frac{1}{2\pi}
\frac{1}{2s\bar{M}^{2} }\text{e}^{-\frac{x^2}{2s\bar{M}^{2} }} \left[
I_{|m-\Phi|}\left(\frac{x^{2} }{2s\bar{M}^{2} }\right)
-I_{|m|}\left(\frac{x^{2} }{2s\bar{M}^{2} }\right)\right]\,.
\end{equation}
By substituting
$t=\frac{x^2}{2s\bar{M}^{2} } = \frac{\rho^{2} }{2s} $, one arrives at
\begin{eqnarray*}
e_+&=&-2\int\limits^0_{t_{max}} \ud t\,\sqrt{\frac{2t}{\rho^2}}
\frac{\ud}{\ud t} \frac{1}{2\pi}\sqrt{\frac{2t}{\rho^2}}
\sum\limits^\infty_{m=-\infty}\frac{1}{2\pi}\frac{t}{\rho^2}\text{e}^{-t}
\left[ I_{|m-\Phi|}(t)-I_{|m|}(t)\right]\\
&=&\frac{2}{2\pi^2\rho^4}\int\limits^{t_{max}}_0 \ud t\,t^{1/2}
\frac{\ud}{\ud t}t^{3/2}
\text{e}^{-t}\left(I_\Phi(t)-I_0(t)+\sum\limits^\infty_{m=1}
\left[ I_{m-\Phi}(t)+I_{m+\Phi}(t)-2I_m(t)\right]\right)\,.
\end{eqnarray*}
After inserting the integral representation \cite{Grad}
\[I_\nu(t)=\frac{1}{\pi}\int\limits^\pi_0\text{e}^{t\cos(\theta)}
\cos(\nu\theta)
\ud\theta-\frac{\sin(\nu\pi)}{\pi}\int\limits^\infty_0
\text{e}^{-t\cosh(z)-\nu z}\ud z\]
into $e_+$, the first terms of the integral representations of the
different modified Bessel functions cancel, so that the only
non-vanishing contribution reads
\begin{eqnarray*}
e_+&=&-\frac{2}{2\pi^2 \rho^4}
\int\limits^{t_{max}}_0\ud t\,t^{1/2}\frac{\ud}{\ud t}
t^{3/2}\text{e}^{-t}\frac{1}{\pi}\left(\int\limits^\infty_0dz\,
\text{e}^{-t\cosh(z)}\sin(\Phi\pi)\text{e}^{-\Phi z}\right.\\
&+&\left.\sum\limits^\infty_{m=1}\int\limits^\infty_0\ud z\,
\text{e}^{-t\cosh(z)}\left(-\cos(m\pi)\sin(\Phi\pi)
\text{e}^{-(m-\Phi)z}+\cos(m\pi)sin(\Phi\pi)
\text{e}^{-(m+\Phi)z}\right)\right)\\
&=&-\frac{2}{2\pi^2 \rho^4}\frac{\sin(\Phi\pi)}{\pi}
\int\limits^{t_{max}}_0\ud t\,t^{1/2}\frac{d}{dt}t^{3/2}
\int\limits^\infty_0\ud z\,\text{e}^{-t(\cosh(z)+1)}
\frac{\cosh(z[\frac{1}{2}-\Phi])}{\cosh(\frac{z}{2})}\,.
\end{eqnarray*}
Partially integrating with respect to $t$ and noting that the surface term
vanishes in the limit $t_{\text{max}}\rightarrow\infty$, one obtains
\[e_+=\frac{2}{(2\pi\rho^2)^2}\frac{\sin(\Phi\pi)}{\pi}
\int\limits^{t_{max}}_0\ud t\,t\int\limits^\infty_0\ud z\,
\text{e}^{-t(\cosh(z)+1)}
\frac{\cosh(z[\frac{1}{2}-\Phi])}{\cosh(\frac{z}{2})}\,.\]
For $\rho>0$ the limit $s_{\text{min}}\rightarrow 0$ implies 
$ t_{\text{max}}\rightarrow\infty$. Then the integration over $t$ can be
carried out with a finite result,
\begin{eqnarray*}
e_+&=&\frac{2}{(2\pi \rho^2)^2}\frac{\sin(\Phi\pi)}{\pi}
\int\limits^\infty_0\ud z\,\frac{1}{(\cosh(z)+1)^2}
\frac{\cosh(z[\frac{1}{2}-\Phi])}{\cosh(\frac{z}{2})}\\
&=&\frac{2}{(2\pi \rho^2)^2}\frac{\sin(\Phi\pi)}{\pi}
4\int\limits^\infty_0\ud z\,\frac{\text{e}^{z(1/2-\Phi)}
+\text{e}^{-z(1/2-\Phi)}}{(\text{e}^{z/2}+\text{e}^{-z/2})
(\text{e}^z+\text{e}^{-z}+2)^2}\,.
\end{eqnarray*}
Substituting $x=\text{e}^z$ yields
\begin{eqnarray*}
e_+&=&\frac{2}{(\pi \rho^2)^2}\frac{\sin(\Phi\pi)}{\pi}
\int\limits^\infty_1\frac{\ud x}{x}\,
\frac{x^{1/2-\Phi}+x^{\Phi-1/2}}{(x^{1/2}+x^{-1/2})(x+x^{-1}+2)^2}\\
&=&\frac{2}{(\pi \rho^2)^2}\frac{\sin(\Phi\pi)}{\pi}
\int\limits^\infty_1\ud x\,\frac{x^{2-\Phi}+x^{\Phi+1}}{(x+1)^5}\\
&=&\frac{2}{(\pi \rho^2)^2}\frac{\sin(\Phi\pi)}{\pi}
\frac{\Gamma(3-\Phi)\Gamma(2+\Phi)}{\Gamma(5)}\\
&=&\frac{1}{\rho^4}\frac{1}{12\pi^2}\Phi(1-\Phi^2)(2-\Phi)\ 
\text{ for }\ \Phi\in(0,1)\,.
\end{eqnarray*}
For $e_-$, i.e. $\lambda(a)=-1$, one obtains the same result in a completely
analogous fashion. Furthermore, according to (\ref{vortexenergie1}) the energy
density is invariant under integer shifts in $\Phi $. The total
energy density  
\begin{equation}
e_{\text{ges}}=e_++e_-
=\frac{1}{\rho^4}\frac{1}{6\pi^2}\phi(1-\phi^2)(2-\phi)\, \ \
\text{for } \ \
\Phi=N+\phi,\,N\in\mathbbm{N}_0,\,\phi\in[0,1)
\label{energievortex}
\end{equation}
\begin{figure}
\begin{center}
\hspace{-1cm}
\epsfig{file = 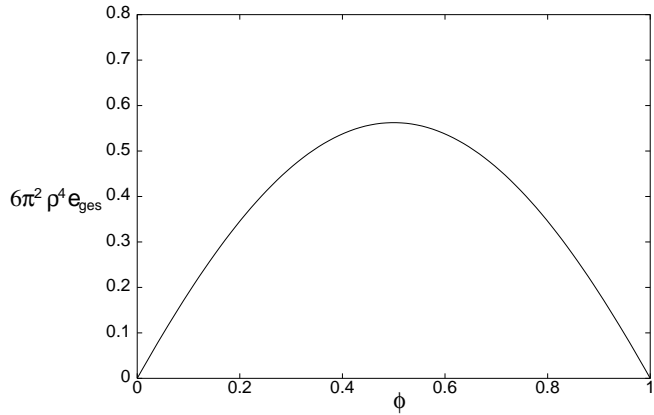, width = 9cm}
\end{center}
\caption{\label{energiedichte2} Energy density $e_{\text{ges}}$ as a
function of magnetic flux $\phi$.}
\end{figure}
is depicted in Fig.~\ref{energiedichte2}\footnote{This result agrees with the
one obtained in ref. \cite{R3}. However, there the freedom in the choice of
the self-adjoint extension was not discussed and a particular choice was
made ad hoc.}. The fact that the energy
density is proportional to $\frac{1}{\rho^4}$ is natural on dimensional
grounds. Furthermore, for integer-valued $\Phi$, the fluctuations are
equivalent to the ones in the vacuum, as expected for background fields
representing pure
gauges in the theory without quarks, i.e. in the presence 
of fluctuating fields in the adjoint representation of the gauge group only.
Such fluctuations are not sensitive to a thin integer flux vortex.

\section{Fermions in the vortex background field}
\label{sec3}

In this section we calculate the energy density of the quark fluctuations
in a vortex background field. The Hamiltonian of the quarks $ \psi (x)$ in
the gauge field $A_\mu (x)$ satisfying the Weyl gauge $A_0 (x) =0$ is
given by 
\begin{eqnarray}
H = \int \ud^3 x \psi ^\dagger (\vec x) h (\vec x) \psi (\vec x) 
\end{eqnarray}
where 

\begin{eqnarray}
\label{G38}h(\vec x) = \vec \alpha \lk \vec p + g \vec A \rk + \beta m
\end{eqnarray}
is the Dirac Hamiltonian with $\vec \alpha , \beta$ denoting the usual
Dirac matrices. In the following we will,
for simplicity, consider massless quarks $\lk m = 0 \rk$, which is a good
approximation for light quark flavours. Furthermore, we will use the Weyl
representation of the Dirac matrices, such that the Dirac operator, eq.
(\ref{G38}), becomes block diagonal,

\begin{equation}
\label{G39}h =\left(\begin{array}{cc}
D & 0 \\ 0 & -D\end{array}\right),\qquad D = \vec\sigma \lk  \vec p +
g \vec A \rk
\end{equation}
where $\vec \sigma$ are the usual Pauli spin matrices. For the gauge potential
of the vortex field defined by eq. (\ref{G1}) the operator
(\ref{G39}) is
diagonal in color space 

\begin{eqnarray}
\label{G40}D_{ab} = \delta _{ab} D (a) , \qquad D (a) =  \vec \sigma \lk 
- \im  \vec \nabla + \frac{\lambda (a)}{2} \frac{\Phi}{\rho} \vec e
_\varphi \rk
\end{eqnarray}
where $\lambda (a) = 1, -1 $ for $a = 1, 2$. Due to the geometry of the
vortex field,
it is convenient to use cylindrical coordinates $(\rho,\varphi,z)$ in which
the operator, eq. (\ref{G40}),
reads

\begin{eqnarray}
D(a)=
\begin{pmatrix}
\label{G41} - \im \partial _z &  \im \text{e}^{-\im\varphi} \lk -
\partial _\rho -
\frac{1}{\rho} \lk l+ \frac{\lambda (a)}{2} \Phi \rk \rk  \\
- \im \text{e}^{\im\varphi} \lk \partial _\rho - \frac{1}{\rho} \lk l +
\frac{\lambda
(a)}{2} \Phi \rk \rk & \im \partial _z
\end{pmatrix}
\end{eqnarray}
where $l = -\im \partial _\varphi$ is the orbital angular momentum. It
is seen that the vortex field generates an intrinsic angular momentum
(color spin) of
the quarks

\begin{eqnarray}
s =\frac{1}{2} \lambda (a) \Phi .
\end{eqnarray}
For center vortices, where $\Phi = 1$, and, since $|\lambda (a)| = 1$,
the total (grand) 
angular momentum of the fermions $J=l+s$ becomes half-integer valued, which is
reminiscent of a boson-fermion transmutation. 

To find the spinor eigenfunctions $u$ of the operator $D(a)$,
cf.~(\ref{G41}), we make the following ansatz,

\begin{eqnarray}
\label{G43}u(a,k,m,N';\vec x)=\text{e}^{\im kz}\begin{pmatrix}\text{e}^{\im
m\varphi}\chi(a,k,m,N';\rho)\\\text{e}^{\im(m+1)\varphi}
\psi(a,k,m,N';\rho)\end{pmatrix},\,k\in\bR,\,m\in\bZ
\end{eqnarray}
where the quantum number $N'$ will be specified further below. With this
ansatz, the eigenvalue equation for the operator $D(a)$, eq. (\ref{G41}),
reduces to 

\begin{eqnarray}
\label{G44}\begin{pmatrix}k&\im(-\partial_\rho-\frac{1+\nu}{\rho})\\
-\im(\partial_\rho-\frac{\nu}{\rho})&-k\end{pmatrix}\begin{pmatrix}
\chi(a,k,m,N';\rho)\\\psi(a,k,m,N';\rho)\end{pmatrix}=E(a,k,m,N')
\begin{pmatrix}\chi(a,k,m,N';\rho)\\\psi(a,k,m,N';\rho)\end{pmatrix}
\end{eqnarray}
where we have introduced the quantum number
\begin{equation}
\nu:=m+\frac{\lambda(a)\Phi}{2}
\end{equation}
of the grand angular momentum $J=l+s$. Note that the eigenvalue equation
(\ref{G44}) is
equivalent to the Dirac equation in two space-time dimensions for a particle
with mass $k$.

In the standard fashion, eq.~(\ref{G44}) can be reduced to a second order
differential equation for the radial functions $\chi, \psi$ which takes
the form of Bessel's differential equation. The solutions of the eigenvalue
eq. (\ref{G44}) for $E^2 > k^2$ (scattering states) are given by

\begin{equation}
\begin{array}{lcl}
\label{G45}\begin{pmatrix}\chi\\\psi\end{pmatrix} & = & \mathcal{N}
\begin{pmatrix}\sqrt{E+k}(\epsilon)^mJ_{\epsilon\nu}(\kappa\rho)\\
\im\sqrt{E-k}(\epsilon)^{m+1}J_{\epsilon(\nu+1)}(\kappa\rho)
\end{pmatrix}\;, \qquad \text{for } E > 0\\
\begin{pmatrix}\chi\\\psi\end{pmatrix} & = & \mathcal{N}
\begin{pmatrix}\im \sqrt{-E-k}(\epsilon)^mJ_{\epsilon\nu}(\kappa\rho)\\
\sqrt{-E+k}(\epsilon)^{m+1}J_{\epsilon(\nu+1)}(\kappa\rho) \end{pmatrix}
\;, \qquad
\text{for } E<0
\end{array}
\end{equation}
where $\mathcal{N}$ is a normalization constant and
$\kappa = \sqrt{E^2 - k^2}$ is
the wave number in the plane perpendicular to the center vortex. Furthermore,
$J_\alpha (x) $ denotes the ordinary Bessel functions and we have introduced
the parameter $\epsilon =\pm 1$. The wave functions, eq. (\ref{G45}), are
solutions of eq.
(\ref{G44}) for both values of this parameter and for the eigenvalues $E=\pm
\sqrt{\kappa^2 + k^2}$.

For $\nu = n \in \bZ$ the solutions with $\epsilon = \pm 1$ are linearly
dependent, since the Bessel functions satisfy the relations
$J_{-n}(x) = \lk -1 \rk ^n J_n (x)$; thus, one should take Bessel functions
and Neumann functions as linearly independent solutions. For
$\nu \notin \bZ$, which holds in particular for a center vortex field, for
which $\Phi = 1$, the solutions with $\epsilon = \pm 1$ are linearly
independent, i.e.~there are two linearly independent solutions for each
eigenvalue. However, requiring the
functions to be square integrable on the $z$-axis selects out a single
solution (that is, either
$\epsilon = 1$ or $\epsilon =-1$) except for $-1<\nu <0$. The latter
interval includes the case $\nu = -\frac{1}{2}$
which occurs for a center vortex, $\Phi = 1$ (when $\lambda (a) = 1$
and $m = -1$ or when $\lambda (a) = -1$ and $m=0$). In this case, both
solutions $\epsilon = \pm 1$ are square integrable on the $z$-axis.

In the following, we will investigate the case $-1<\nu < 0$
in more detail. The point is that we have to impose
appropriate boundary conditions on the $z$-axis, where the vortex field has a
singularity, so that a unique solution survives. The boundary conditions
have to be chosen in such a way that the operator in eq.~(\ref{G40})
becomes self-adjoint. The procedure works analogously to the gluonic
case: One restricts the domain of definition of the operator $D(a)$,
eq.~(\ref{G40}), in a suitable way, such that it becomes symmetric. Then
one determines the deficiency spaces of this operator and,
from this, all its self-adjoint extensions. The construction of all
self-adjoint extensions of the operator in eq.~(\ref{G44}) was given in
\cite{Gerbert,Falomir}. The
extensions are characterized by a parameter $\theta \in [ 0, 2\pi ) $ and
defined by the following boundary conditions on the eigenfunctions,

\begin{eqnarray}
\label{G46}\begin{pmatrix}\chi\\\psi\end{pmatrix}\rightarrow
\begin{pmatrix}\im(\bar{M} \rho)^\nu\sin\left(\frac{\pi}{4}+
\frac{\theta}{2}\right)\\(\bar{M} \rho)^{-\nu-1}\cos\left(\frac{\pi}{4}+
\frac{\theta}{2}\right)\end{pmatrix}\qquad\text{for }\rho\rightarrow 0,
\end{eqnarray}
where $\bar{M} $ is a mass scale introduced in order to have dimensionless
quantities like $\bar{M} \rho$. The
requirement of self-adjointness restricts the choice of the boundary 
conditions
on the vortex axis ($z$-axis). For the boundary conditions (\ref{G46}), a
unique solution follows for each value of $\theta \in [ 0, 2\pi ) $,
which in the case of $E>0$ is given by 

\begin{eqnarray}
\begin{pmatrix}\chi\\\psi\end{pmatrix}=\mathcal{N}[1+(-1)^m\sin(2\mu)
\cos(\nu\pi)]^{-1/2}\nonumber \\
\begin{pmatrix}\sqrt{E+k}[\sin(\mu)J_\nu(\kappa\rho)+
(-1)^m\cos(\mu)J_{-\nu}(\kappa\rho)]\\\im\sqrt{E-k}[\sin(\mu)J_{\nu+1}
(\kappa\rho)+(-1)^{m+1}\cos(\mu)J_{-(\nu+1)}(\kappa\rho)]\end{pmatrix}
\end{eqnarray}
where the quantity $\mu$ is related to $\theta$ by

\begin{eqnarray}
\tan\left(\frac{\pi}{4}+\frac{\theta}{2}\right)=
(-1)^m\left(\frac{E+k}{E-k}\right)^{1/2}\left(\frac{\kappa}{2k}\right)^{2\nu+1}
\frac{\Gamma(-\nu)}{\Gamma(\nu+1)}\tan(\mu).
\end{eqnarray}
In addition to these scattering states, for $\frac{\pi}{2}< \theta
<\frac{3\pi}{2}$ bound states, i.e. square integrable eigenstates exist
\cite{Gerbert,Falomir}.

Finally, let us remark that for $\nu < -1$ and for $\nu >0$, where the
parameter $\epsilon = \pm 1$ is already uniquely determined by requiring
square integrability of the wave function on the $z$-axis, the
application of the above procedure for finding self-adjoint 
extensions of the Dirac operator is trivial. In fact, one finds that the
deficiency spaces of this operator are empty and that hence this operator is
essentially self-adjoint. 

In those cases where there is more than one self-adjoint extension, one has
to select one on physical grounds.
In reference \cite{Manuel} it was shown that only for
$\theta = \pm \frac{\pi}{2}$ a vortex on the $z$-axis can be consistently
described in the sense that the eigenfunctions of the self-adjoint
extensions of the Dirac operator and its square fulfill the same
boundary conditions. It is still
open which sign of $\theta$ has to be chosen for each flux
$\Phi$.
The sign of $\theta$ can be fixed by requiring that
the energy of the quarks in a vortex field should be a periodic function
in the flux $\Phi$ of period 2.
This is because fermions should not feel a Dirac string, $\Phi = 2$.
It turns out that this requires choosing $\theta $ such that the assignments
listed in the table below, cf.~(\ref{tabel}), result.
Let us emphasize that this implies different signs for different ranges of
flux. The same self-adjoint extension of the Dirac operator was
obtained in \cite{Sitenko} by demanding that the expectation value of
the resulting baryon number

\begin{eqnarray}
\langle N\rangle=-\frac{1}{2}\sum_n \text{sgn}(E_n)
\end{eqnarray}
properly transforms under charge conjugation,
$\langle N \rangle \rightarrow - \langle N \rangle$,
and is a periodic function in the flux $\Phi$ with period $2$.

It turns out that this choice of the parameter $\theta$ corresponds to 
choosing the minimal irregularity of the eigenfunctions on the vortex axis.

\begin{equation}
\begin{array}{|c|l|l|}\hline
&0<\Phi<1&1<\Phi<2\\\hline
\lambda(a)=1&\epsilon=1\text{ for }m=0,1,... &\epsilon=1\text{ for }m=-1,0,... \\
&\epsilon=-1\text{ for }m=-1,-2,... &\epsilon=-1\text{ for }m=-2,-3,... \\\hline
\lambda(a)=-1&\epsilon=1\text{ for }m=0,1,... &\epsilon=1\text{ for }m=1,2,...\\
&\epsilon=-1\text{ for }m=-1,-2,... &\epsilon=-1\text{ for }
m=0,-1,...\\\hline 
\end{array} \label{tabel}
\end{equation}
Furthermore, the same choice of self-adjoint extension was obtained in
reference \cite{Beneventano} by choosing boundary conditions of the
Atiyah-Patodi-Singer type for a finite radius of the vortex flux and
subsequently letting this radius go to zero.

\subsection{The energy density of quarks in a vortex background}

In the following, we calculate the energy density of fermion fluctuations
in the vortex gauge field. The energy density is defined by 

\begin{eqnarray}
e(\boldsymbol{x})=\sum\limits_{a=1,2}\sum\limits_
{\tilde{E} (a,N,n)<0}\tilde{E} (a,N,n)(\Psi^a)
^\dagger(a,N,n;\boldsymbol{x})\Psi^a(a,N,n;\boldsymbol{x})
\end{eqnarray}
where $\tilde{E}$ and $\Psi$ denote the energy eigenvalues and
eigenfunctions of the Dirac Hamiltonian, eq.~(\ref{G38}). Due to the
block structure of the Dirac Hamiltonian, these eigenvalues are given
by both the positive and negative eigenvalues $E$ appearing in 
eq.~(\ref{G44}), and the energy density becomes 

\begin{equation}
\begin{array}{lcl}
e(\boldsymbol{x}) & = & \sum\limits_{a=1,2} \Big(
\underbrace{\sum\limits_{N,E(a,N)<0}
E(a,N)u^\dagger(a,N;\boldsymbol{x})u(a,N;\boldsymbol{x})}
_{=:e(a,E<0;\boldsymbol{x})} \\
&&  - \underbrace{\sum\limits_{N,E(a,N)>0}
E(a,N)u^\dagger(a,N;\boldsymbol{x})u(a,N;\boldsymbol{x})}
_{=:e(a,E>0;\boldsymbol{x})} \Big)
\end{array}
\end{equation}
with $u$ denoting the wave functions corresponding to the energy eigenvalues
$E$. We have assumed here that the wave functions are properly
normalized (as scattering states),

\begin{eqnarray}
\int_0 ^{\infty} \rho d \rho \hspace{0.3cm}
u^\dagger \lk a, N, \kappa , \vec x \rk u \lk a, N, \kappa ', \vec x \rk
= \frac{1}{\kappa} \delta \lk \kappa - \kappa ^{'} \rk .
\end{eqnarray}
This fixes the normalization constant to 

\begin{eqnarray}
\mathcal{N} = \frac{1}{\sqrt{2|E|}}.
\end{eqnarray}
We have to choose the eigenfunctions such
that they respect the boundary conditions corresponding to the minimal
irregularity on the vortex axis (\ref{tabel}).

With this choice of the eigenfunctions, one finds after a straightforward
calculation the energy density for fluxes $0<\Phi < 1$,

\begin{equation}
\label{57}
e_1(\Phi;\boldsymbol{x})=-\frac{1}{2\pi^2}\int_{-\infty}^\infty
\ud k\,\int_0^\infty\kappa
\ud\kappa\,\sqrt{\kappa^2+k^2}\left(2\sum\limits_{m=-\infty}^\infty
J^2_{|m+\Phi/2|}(\kappa\rho)+J^2_{-\Phi/2}(\kappa\rho)
-J^2_{\Phi/2}(\kappa\rho)\right)
\end{equation}
while for fluxes $1<\Phi < 2$ one obtains  

\begin{equation}
\label{58}
e_2(\Phi;\boldsymbol{x})=-\frac{1}{2\pi^2}\int_{-\infty}^\infty
\ud k\,\int_0^\infty\kappa
\ud\kappa\,\sqrt{\kappa^2+k^2}\left(2\sum\limits_{m=-\infty}^\infty
J^2_{|m+\Phi/2|}(\kappa\rho)+J^2_{\Phi/2-1}(\kappa\rho)
-J^2_{1-\Phi/2}(\kappa\rho)\right).
\end{equation}
These two expressions for the energy density coincide for the flux $\Phi = 1$,
which corresponds to a center vortex. Thus, the energy density as a
function of the flux $\Phi$ is continuous at the center vortex flux
$\Phi = 1$. Furthermore, one also immediately recognizes that the relation
$e_1(\Phi=0;\boldsymbol{x}) =e_2(\Phi=2;\boldsymbol{x})$ holds,
which implies that the energy density of a Dirac string, $\Phi = 2$, is
equal to the energy density of the perturbative vacuum, $\Phi = 0$.
This is expected, since the Dirac string does not carry observable physical
flux. Let us emphasize that this degeneracy in the energy density justifies
the self-adjoint continuation chosen above.

\begin{figure}
\begin{center}
\epsfig{file = 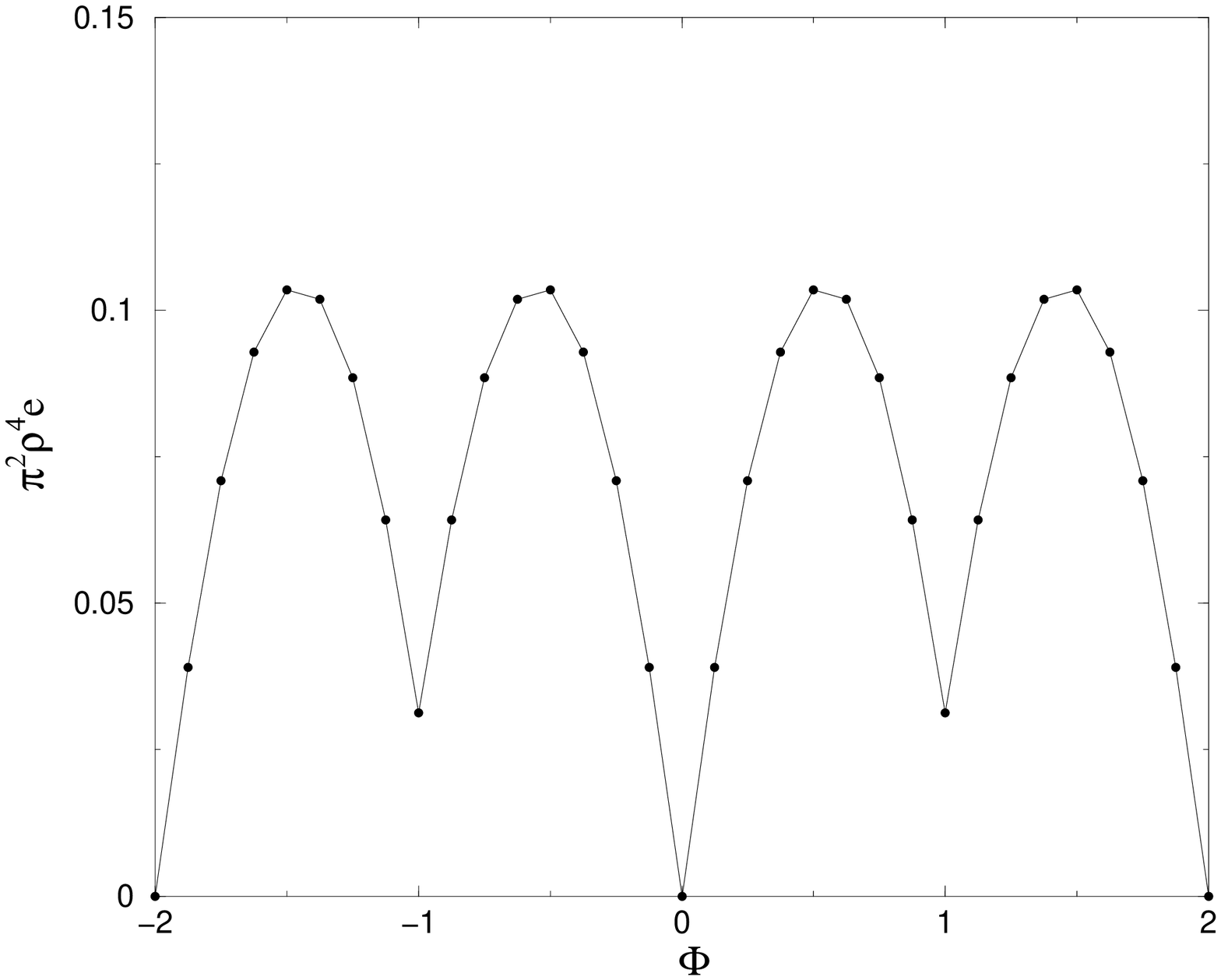, width = 9cm}
\end{center}
\caption{\label{figure2}Total energy density of the fluctuating gauge
field and of the fermions.}
\end{figure}

Since the energy eigenvalues depend only on the combination $\lambda (a) \Phi$ 
and summation is carried out over $\lambda (a) = \pm 1$, the energy density is
reflexion symmetric, i.e. stays invariant under the replacement
$\Phi \rightarrow -\Phi $. Furthermore, as can be read off from the explicit
expressions (\ref{57}), (\ref{58}),
the energy densities for the flux ranges $0<\Phi <1$ and
$1<\Phi <2$ are related by 

\begin{eqnarray}
e_1 \lk \Phi \rk = e_2 \lk 2-\Phi \rk .
\end{eqnarray}
We can therefore restrict ourselves to the flux interval $0<\Phi <1$.
Subtracting from the energy density of the quark fluctuations in the
presence of the vortex the corresponding energy density in the perturbative
vacuum, one finds

\begin{eqnarray}
\Delta
e_1(\Phi; \boldsymbol{x}) & = &e_1(\Phi;\boldsymbol{x})-e_1 (\Phi = 0;
\boldsymbol{x})\\
& = & \underbrace{-\frac{2}{2\pi^2} \int \limits_{-\infty}^\infty \ud k\,
\int \limits_0^\infty \kappa
\ud\kappa\,\sqrt{\kappa^2+k^2} \sum \limits_{-\infty}^{\infty}
\left(J^2_{|m+\Phi /2|}(\kappa\rho) - J^2_{|m|}(\kappa\rho) \right)}_{=:A}\\
&& \underbrace{-\frac{1}{2\pi^2} \int \limits_{-\infty}^\infty \ud k\, \int
\limits_0^\infty \kappa
\ud\kappa\,\sqrt{\kappa^2+k^2}
\left(J^2_{-\Phi/2}(\kappa\rho)-J^2_{\Phi/2} (\kappa\rho)\right)}_{=:B}
\end{eqnarray}
The sums which occur here are identical to the sums which occur in the gluonic
part of the energy density. In fact, using the results obtained for the
gluonic part, see eq.~(\ref{vortexenergie1}), and replacing $\Phi$ by
$\Phi /2$, one finds for the first term
%in the excess energy density of
%the quark fluctuations in the presence of the vortex field
\begin{eqnarray}
A=-\frac{1}{6\pi^2\rho^4}\frac{\Phi}{2}
\left(1-\left(\frac{\Phi}{2}\right)^2\right)
\left(2-\frac{\Phi}{2}\right).
\end{eqnarray}
Analogous calculations as in the gluon sector yield for the second term 

\begin{eqnarray}
B=-\frac{1}{\pi^2\rho^4}\int\limits_0^\infty 
\ud t\,t^{1/2}\frac{\ud}{\ud t}t^{3/2}\text{e}^{-t}\left
(I_{-\Phi/2}(t)-I_{\Phi/2}(t)\right).
\end{eqnarray}
The remaining integral has to be carried out numerically for given vortex flux
$\Phi$. Adding this contribution to the gluonic part, we obtain the
total energy
density of the fluctuating gauge field and of the fermions in the vortex
background field. The result is shown in Fig.~\ref{figure2}.

\section{Discussion}
\label{sec4}
We have evaluated the one-loop effective energy density associated with an
infinitely thin SU(2) Yang-Mills vortex, taking into account both gluonic
as well as (massless) quark fluctuations, for arbitrary vortex flux $\Phi $.
As a function of the latter, center vortices (odd integer $\Phi $) represent
local minima of the effective energy -- it is energetically favorable to keep
magnetic flux bundled in units corresponding to center elements of the gauge
group as opposed to letting magnetic flux disassociate and propagate in
lines of arbitrary flux through space. In this sense, the quantum fluctuations
tend to stabilize center vortices with respect to local deformations.

Globally, the thin center vortices we consider are degenerate with the
perturbative vacuum if one takes into account only gluonic fluctuations,
cf.~Fig.~\ref{energiedichte2}. This is natural, since these infinitely
thin, odd integer $\Phi $ configurations correspond to pure gauge 
transformations in pure Yang-Mills theory -- they are just as invisible
to gluonic fluctuations as Dirac strings (even integer $\Phi $), since
gluons are defined in the adjoint representation of the gauge group.
This degeneracy is lifted when quark fluctuations are included. The energy
density of a center vortex is raised above the energy density of the
perturbative vacuum through the influence of the fermions (but it continues
to represent a local minimum as a function of magnetic flux $\Phi $).
Only Dirac strings are then still degenerate with the perturbative vacuum.

In a realistic vortex model of the QCD vacuum, energy densities of the type
considered here are but one among a whole set of factors influencing vortex
dynamics. For example, realistic vortices are not infinitely thin, but are
smeared out to a finite transverse thickness, a case which has recently also
been discussed \cite{maul,bordag}. As shown there, a finite thickness can
actually lead to a lowering of the effective energy density below the one
of the perturbative vacuum. On the other hand, also vortex interactions
have to be considered, cf.~\cite{R4} for an investigation on the lattice.
In the phenomenologically successful random vortex world-surface model
\cite{rand,preptop,csb}, the energy associated with vortex curvature was
found to play an important role -- this contribution has not been investigated
from first principles. In that model, it is ultimately the entropy inherent
in the random vortex world-surface ensemble which overcomes any energy
penalties associated with the formation of center vortices, and thus leads
to the formation of the percolating vortex vacuum and a comprehensive
description of the principal nonperturbative phenomena characterizing the
strong interaction.

\begin{acknowledgments}
We thank P.A.G.~Pisani for helpful discussions, especially about the proof
given in Appendix~\ref{append1}.
\end{acknowledgments}

\begin{appendix}
\section{Closure of $\hat{C}$} \label{append1}
In this Appendix we discuss the closure of the operator $\hat{C}$,
cf.~(\ref{chat}). We prove
that functions $f \in D(\overline{\hat{C}})$ are absolute continuous
close to the origin  and
fulfill $\lim_{x \rightarrow 0} x^{-\delta} f(x) = 0$, where $\delta =
|m-\lambda(a)\Phi|$. To obtain $D(\overline{\hat{C}})$, we add to the domain
of $\hat{C}$, $D(\hat{C})= C^\infty_c (\mathbbm{R}_{>0})$, the limit
points of Cauchy sequences in $D(\hat{C})$, whose images under $\hat{C}$
are also Cauchy sequences. We consider such a Cauchy sequence $f_k$ and
denote $f_{kl} = f_k -f_l$. We obtain
\begin{eqnarray*}
\int_0^\infty \ud x\; (x {f_{kl}'}^2 + \frac{\delta}{x} {f_{kl}}^2) & = &
\int_0^\infty x\; \ud x\; f_{kl} \hat{C} f_{kl} < \epsilon\;, \text{ for } k,l >
N_\epsilon \;.
\end{eqnarray*}
Therefore, $x^{-1/2} f_k$ and $x^{1/2} f_k'$ are Cauchy sequences in
$L^2(\mathbbm{R}_{>0} , \ud x)$. Observe that also $x^{1/2} f_k$ and
$x^{1/2} \hat{C} f_k$ are Cauchy sequences in $L^2(\mathbbm{R}_{>0},
\ud x)$.

Next we prove that if $x^n f_k'$ and $x^{n-1} f_k$ are Cauchy sequences
($n\in\mathbbm{R}$) in $L^2(\mathbbm{R}_{>0}, \ud x)$,
then also $x^{n/2-1/4} f_k'$ and $x^{n/2-5/4} f_k$ are Cauchy
sequences. From that, it follows by induction that $x^{-1/2+\epsilon} f_k'$
and $x^{-3/2 + \epsilon}f_k$ are Cauchy sequences for any $0 < \epsilon
\ll 1$. We consider $\int_0^\infty \ud x\; x^n f_{kl}'
\sqrt{x} \hat{C} f_{kl} < \xi \text{ for } k,l > N_\xi$
and obtain by partial integration
\begin{eqnarray*}
\int_0^\infty \ud x\; \left( (\frac{n}{2} - \frac{3}{4})x^{n-1/2}
{f_{kl}'}^2 - \frac{\delta}{2} (n-\frac{3}{2}) x^{n-5/2} {f_{kl}}^2
\right) & < & \xi \;.
\end{eqnarray*}
We obtain another relation from
$\int_0^\infty \ud x\; x^{n-1} f_{kl} \sqrt{x} \hat{C} f_{kl} <
\xi \;, \text{ for } k,l > N_\xi$:
\begin{eqnarray*}
\int_0^\infty \ud x\; \left( x^{n-1/2} {f_{kl}'}^2 + (\delta -
\frac{(n-3/2)^2}{2} ) x^{n-5/2} {f_{kl}}^2 \right) & < & \xi \;.
\end{eqnarray*}
For $\delta \notin \mathbbm{Q}$, $n\in \mathbbm{Q}$ and vice versa,
which can be achieved for any $0 \leq \delta <1$ by adding an
infinitesimal number to $n$, these two relations are linearly
independent, as one can check by calculating the determinant. In these
cases $x^{n/2-5/4} f_{kl}$ and $x^{n/2-1/4} f_{kl}'$ are Cauchy
sequences in $L^2(\mathbbm{R}_{>0}, \ud x)$.

In the next step, we prove that the functions in $D(\overline{\hat{C}})$
are absolute continuous close to the origin. We consider an interval
$(0,a)$ and denote the limit of $\sqrt{x} f_k$ by $f$, the
limit of
$\sqrt{x}\hat{C}f_k$ by $f_{\hat{C}}$, the limit of $x^{-1/2+\epsilon}
f_k'$ by $g$ and the limit of $x^{-3/2+\epsilon} f_k$ by $\tilde{g}$ in
$L^2(\mathbbm{R}_{>0},\ud x)$. We get
\begin{eqnarray*}
\int_0^a \ud x\; |f_k-x^{3/2-\epsilon} \tilde{g} |^2 & \leq &
a^{3-2\epsilon} \int_0^a \ud x\; |x^{-3/2 + \epsilon}f_k -
\tilde{g} |^2 < \xi 
\end{eqnarray*}
that is
\begin{eqnarray*}
f_k & \rightarrow x^{3/2-\epsilon} \tilde{g} \equiv h
\end{eqnarray*}
in $L^2((0,a),\ud x)$. On the other hand, as $g \in L^2((0,a),\ud
x)$ and therefore the integral $\int_0^x \ud x'\; x'^{1/2-\epsilon} g$
is well defined for all $x\in(0,a)$, we have
\begin{eqnarray*}
\left|\int_0^x \ud x' \; x'^{1/2-\epsilon} g - f_k \right|^2 & \leq &
a^{1-2\epsilon} \int_0^x \ud x' \; | g - x^{-1/2+\epsilon} f_k' |^2 <
\xi
\end{eqnarray*}
and see that
\begin{eqnarray*}
f_k & \rightarrow & \int_0^x \ud x'\; x'^{1/2-\epsilon} g \;.
\end{eqnarray*}
However, as $f_k \rightarrow h$ we see that $h$ is absolute continuous in
$(0,a)$ and $\frac{h}{x^{1-2\epsilon}}$ is absolute
continuous in $(a',a)$ for $0<a'<a$. Taking the derivative of
$\frac{h}{x^{1-2\epsilon}}$ we obtain
\begin{eqnarray*}
(\frac{h}{x^{1-2\epsilon}})' & = & g \frac{1}{x^{1/2-\epsilon}} -
(1-2\epsilon) \tilde{g} \frac{1}{x^{1/2-\epsilon}} \in
L^2((0,a),\ud x)
\end{eqnarray*}
as all functions on the
right side are in $L^2((0,a),\ud x)$. Therefore
\begin{eqnarray*}
\int_0^a \ud x\; (\frac{h}{x^{1-2\epsilon}})' & = &
\frac{h(a)}{a^{1-2\epsilon}} - \lim_{a'\rightarrow
0} \frac{h(a')}{{a'}^{1-2\epsilon}}
\end{eqnarray*}
has to be finite and we arrive at
\begin{eqnarray*}
\lim_{a'\rightarrow 0} h(a')
{a'}^{-1+2\epsilon} & = & \text{const} < \infty
\end{eqnarray*}
for any $0<\epsilon \ll 1$, i.e.
\begin{eqnarray*}
x^{-\delta} h(x) & \rightarrow & 0 \text{ for } x \rightarrow 0
\end{eqnarray*}
for all $0 \leq \delta <1$.
\end{appendix}

\end{document}